# PowerMat: context-aware recommender system without user item rating values that solves the cold-start problem

**Hao Wang**

CEO Office, Ratidar.com, Beijing, 100080, China

haow85@live.com

Abstract—Recommender systems serves as an important technical asset in many modern companies. With the increasing demand for higher precision of the technology, more and more research and investment has been allocated to the field. One important sub-field of recommender systems that has been stagnating is context-aware recommender systems. Due to the difficulty of collecting input dataset, the amount of research on context-aware recommender systems is much less than other sub-fields of recommender systems. In this paper, we propose a new algorithm named PowerMat to tackle the context-aware recommendation problem. We build our theory on matrix factorization and Zipf's law, and also more recent research work such as DotMat. We prove by experiments that our method achieves superior results to the classic matrix factorization algorithm and other context-aware recommender systems such as MovieMat+. In addition, by theoretical analysis, we show that our algorithm solves the cold-start problem for context-aware recommendation.

## 1. Introduction

Recommender system is one of the symbolic technologies of Web 2.0. The early adopters of recommender systems believe the old way for finding useful information is search engine, and the new way for information seeking is recommender systems. Following this thought, many corporations such as Baidu, Alibaba and Tencent started to build their own teams for the technology and related products. In the next wave of IT revolutions in the following years, startups such as Toutiao.com and TikTok quickly emerged as major players of the industry with the help of recommendation technology.

In spite of the growing investment and research on the technology, recommender systems have some sub-fields that have been overlooked compared with the mainstream focus on the algorithmic precision. One such sub-field is context-aware recommendation - the field that combines contextual information together with user item rating values to predict the preferences of users over a set of items. A workshop named CARS [1][2][3] has been held as an affiliating research venue with RecSys since early 2010's on the topic, but the number of accepted papers remained very small in recent years. In the industry, corporations such as Burger King and Ctrip have been using the technology to increase their traffic and sales volumes.

One major difficulties of context aware recommendation research is the collection of input data. Open-sourced datasets are highly limited, and many researchers do not have access to the data sets needed for algorithmic design. An important dataset named LDOS-CoMoDa [4] was released online in 2012 that incorporates contextual information together with user's movie ratings. We use this dataset as the major experiment dataset for our research in this paper.

To build our algorithm, we take advantage of the matrix factorization framework as the backbone of our algorithmic design, and we transform the matrix factorization algorithm into a context-aware recommendation algorithm as inspired by Zipf's law and the DotMat [5] algorithm. In the experiment section, we prove that our approach outperforms the classic matrix factorization and another context-aware recommendation algorithm named MovieMat+ [6].

## 2. Related Work

Recommender systems based on matrix factorization have been extremely popular in the industry [7][8]. In addition to the classic matrix factorization algorithms, new inventions emerged in recent years. Zipf Matrix Factorization [9], MatRec [10] and KL-Mat [11] were proposed to solve the fairness problems. Researchers pick matrix factorization as the benchmark to be modified for various scenarios due to its simplicity and interpretability.

Context-aware recommendation has been researched from the last decade. There has been research that utilizes penalty [12] or weighted linear combinations [13] to tackle the problem. Unlike the common approaches, Wang proposed MatMat algorithm [14] and a practical example MovieMat/MovieMat+ [6] that transform the scalar fitting problem of the classic matrix factorization into a matrix fitting problem.

There are different metrics to compare the performance of a recommender system. Classic ones include MAE, RMSE, nDCG, etc. All of them are used to compare the accuracy of the recommender systems. In recent years, a new metric, e.g., Degree of Matthew Effect [9], was invented and used in many papers. There is also another research that proposes to use extreme value theory [15] to evaluate the fairness of recommender systems.

## 3. PowerMat Formulation

Before proposing our own algorithm, we review the idea of DotMat first. DotMat is an algorithm inspired by RankMat, which dismisses the idea of using exponential family to model the probabilistic distribution in matrix factorization. In particular, the loss function of DotMat is as follows :

$$L = |(U_i^T \cdot V_j)^{U_i^T \cdot v_j} - \frac{R_{i,j}}{R_{max}}| \qquad (1)$$

To reconstruct the unknown user item rating values, we use the following formula :

$$R_{i,j} = R_{max} \times (U_i^T \cdot V_j) \qquad (2)$$

The reconstruction formula might appear wrong at the first sight, since in the loss function, we are using the power of $U_i^T \cdot V_j$ to estimate the user item rating value. However, the researcher designed the algorithm in such a way on purpose. By formulating the algorithm in this way, the most popularity items generate the smallest error in the loss function, so the overall performance of the system will be greatly enhanced compared with the original formulation of matrix factorization.

We formulate the PowerMat algorithm as an MAP problem (as probabilistic matrix factorization) :

$$P(U, V \mid R, c) = P(R, c \mid U, V)P(U)P(V) \qquad (3)$$

where R is the user item rating value, c is the contextual information vector, U is the user feature vector, and V is the item feature vector.

We define the posterior distribution P(U, V | R, c) as follows :

$$P(U, V \mid R, c) = (U \cdot V)^{\alpha \cdot c + \beta U \cdot V} \quad (4)$$

where α is the coefficient vector for the contextual information vector, and β is the scalar coefficient for the dot product of U and V. Just like probabilistic matrix factorization, we use normal distributions to model the prior distributions of U and V. In the end, we take the natural log of the loss function, and compute the stochastic gradient descent update formulas for the loss function :

$$U = U - \gamma(\beta U^T \cdot V \cdot V + (\beta U^T \cdot V + \alpha \cdot c)V - \frac{2}{\sigma_U}U) \quad (5)$$
$$V = V - \gamma(\beta V^T \cdot U \cdot U + (\beta V^T \cdot U + c^T \cdot \alpha)U - \frac{2}{\sigma_V}V) \quad (6)$$
$$\alpha = \alpha - \gamma(U^T \cdot V \times c) \quad (7)$$
$$\beta = \beta - \gamma((U^T \cdot V)^2) \quad (8)$$

By closely examining formulas (5) to (8), we observe that to execute the PowerMat algorithm, the only input data that we need is the contextual information, there is no user-item rating values needed in the formula. Our algorithm can actually solve the cold-start problem of context-aware recommendation. For example, when a new user enters a movie theater, we do not know his preferences, but we have access to contextual information thanks to the sensors deployed in the theater room, we would be able to recommend new movies to him on movie theater's apps using PowerMat.

### 4. Experiment

Due to the difficulties to obtain datasets for context aware recommendation, we conduct our experiments on LDOS-CoMoDa datasets. LDOS-CoMoDa is an open datasets that contains multiple contextual information fields. We compare our algorithm against the classic matrix factorization, which is a context-free algorithm, and MovieMat+, which is a context-aware movie recommender system that incorporates 6 different contextual information values into the system.

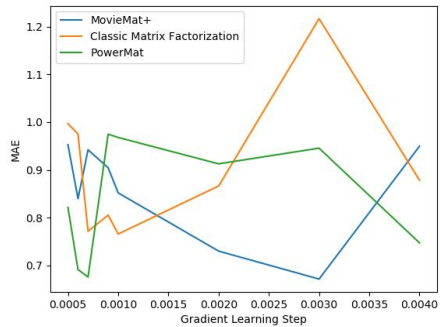

Fig. 1 Comparison of PowerMat
with other algorithms on MAE

Fig. 1 illustrates the accuracy comparison results among PowerMat, classic matrix factorization and MovieMat+. According to the experiment results, PowerMat achieves similar optimal values when the gradient learning step is small, while MovieMat+ outperforms PowerMat in larger parameter values, but never able to reach the optimal value of PowerMat with most parameters. The classic matrix factorization algorithm never outperforms both of the algorithms.

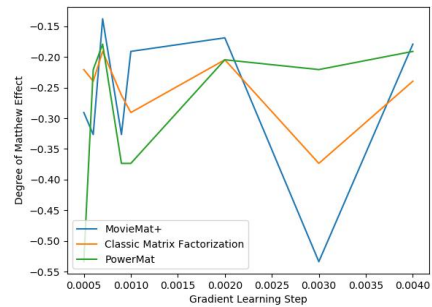

Fig. 2 Comparison of PowerMat with other algorithms on Degree of Matthew Effect

Fig. 2 demonstrates the fairness comparison results among PowerMat, classic matrix factorization and MovieMat+. According to the experiment results, PowerMat achieves optimal fairness values when it achieves the optimal accuracy values and when the parameters are large, while MovieMat+ outperforms PowerMat in the middle section of parameter values. The matrix factorization never outperforms both of the context-aware recommender systems.

## 5. Conclusion

Context-aware recommendation is a long-term research interest among many researchers and industrial engineers. However, since open datasets for context-aware recommendation research are highly limited, the field has not witnessed many disruptive innovations. In this paper, we propose a new context-aware recommender system named PowerMat, that produces competitive results and solve the cold-start problem.

In future, we would like to explore the problem of how to explain the effectiveness of the PowerMat algorithm. This might help us to delve deeper into the knowledge domain of AI.

**Acknowledgments**
I'm grateful to the internet age, of which I'm a beneficiary rather than a victim. Availability of open-sourced literature and cheap computing facilities has helped me through the years of my own independence research. Thanks to the research methodology basics and self-confidence given to me by my professors and alumni, and also my own professional experience and hard work, I've been able to land 3 best paper award / best oral presentation awards by 2021. These awards enabled me to persist when I was alone, unsupported and suicidal.